\documentclass[12pt,bm]{article}
\usepackage{graphicx}
\usepackage{theorem}
\usepackage{makeidx}
\usepackage{amsmath}
\usepackage{epic}
\usepackage{amscd}
\usepackage{bbm}
\usepackage{xy}
\usepackage{amssymb}
\usepackage{times}
\usepackage{verbatim}
\usepackage{epsfig}
\usepackage{bm}
\usepackage{natbib}

\newif\iffigs
\figstrue
\def\a{{\bm a}}
\def\b{{\bm b}}
\def\c{{\bm c}}
\def\v{{\bm v}}
\def\c{{\bm c}}

\def\k{{\bm k}}

\def\z{{\bm z}}
\def\x{{\bm x}}
\def\y{{\bm y}}

\def\L{{\mathrm L}}

\def\ms1{{\kern .035em s}}

\def\pmb#1{\setbox0=\hbox{#1}%
  \kern-.025em\copy0\kern-\wd0
  \kern.05em\copy0\kern-\wd0
  \kern-.025em\raise.0433em\box0 }
\def\pmB#1{\setbox0=\hbox{#1}%
  \kern-.025em\copy0\kern-\wd0
  \kern.035em\copy0\kern-\wd0
  \kern-.025em\raise.0433em\box0 }

\def\ced#1{\setbox0=\hbox{#1}\ifdim\ht0=1ex \accent'30 #1%
  \else{\ooalign{\unhbox0\crcr\hidewidth\char'30\hidewidth}}\fi}

\makeatletter \def\fps@figure{tp} \makeatother
\iffigs
\fi

\def\drawing #1 #2 #3 {
\begin{center}
\setlength{\unitlength}{1mm}
\begin{picture}(#1,#2)(0,0)
\put(0,0){\framebox(#1,#2){#3}}
\end{picture}
\end{center} }

\begin{document}

{\center
{\it Short title: Lagrangian singularities of steady flow}

\bigskip
{\bf Lagrangian singularities of steady two-dimensional flow}

\bigskip
Walter Pauls$^{a,b,}$\footnote{E-mail: walter.pauls@physik.uni-bielefeld.de}
and
Takeshi Matsumoto$^{a,c,}$\footnote{E-mail:
  takeshi@kyoryu.scphys.kyoto-u.ac.jp}

\bigskip
$^a$Observatoire de la C\^ote d'Azur, CNRS UMR~6202,\\
BP~4229, 06304 Nice Cedex 4, France

\bigskip
$^b$ Fakult\"at f\"ur Physik, Universit\"at Bielefeld, \\
Universit\"atsstra\ss e 
25, 33615 Bielefeld, Germany

\bigskip
$^c$ Dep. Physics, Kyoto University, Katashirakawa Oiwakecho\\
Sakyo-ku, Kyoto 606-8502, Japan

\bigskip
{\it Submitted to Geophysical and Astrophysical Fluid Dynamics}\\
 \centerline{Revised version, May 2004}
}

\begin{abstract}\noindent
The Lagrangian complex-space singularities of the steady Eulerian flow with
stream function $\sin x_1 \cos x_2$ are studied by numerical and analytical
methods. The Lagrangian singular manifold is analytic. Its minimum distance
from the real domain decreases logarithmically at short times and
exponentially at large times.

\end{abstract}

\section{Introduction}
\label{s:intro}

Singularities of the solutions of hydrodynamical problems in the complex space
domain are important, for example, because they allow us to give an objective
definition of the otherwise fuzzy concept of ``smallest scale present in a
flow''. Specifically, if a flow is analytic in the space variable $\x$ for all
$\x$, it can be continued to complex $\z=\x+i\y$ locations where it will
generally have singularities on some complex set $\Sigma$.  The minimum
distance $\delta$ of $\Sigma$ to the real domain, called the width of the
analyticity strip, defines the smallest scale. Indeed, the modulus of the
Fourier transforms of the velocity $\v$ decreases roughly as $e ^{-\delta k}$
at high wavenumbers $k$; thus the mesh needed in numerically simulating such a
flow, e.g. by spectral methods, has to be significantly less than $\delta$
\citep{SSU83,BMONMF83}.
Furthermore, should the
flow ever develop a real singularity at a finite real time $t_\star$, it must
be preceded by complex singularities at a distance $\delta(t)$ which
continuously vanishes at $t_\star$ 
\citep{BardosBenachourZerner76,BenachourII76,BenachourI76}. 
For incompressible flow actual 
measurements of $\delta(t)$, using high-resolution spectral simulations,
indicate that $\delta(t)$ decreases in time, but in a way much tamer than
suggested by the known rigorous estimates. In both two and three dimensions
the behavior of $\delta(t)$ at large times may well be exponential, but the
best estimates are a decreasing double exponential in two dimensions and
finite-time blow-up in three dimensions \citetext{see \citet{FMB03} for
review}.

The standard explanation for this discrepancy is the phenomenon
of nonlinear depletion, whereby the flow is found to organize itself
into structures which have nearly vanishing nonlinearities 
\citep{Frisch95,MajdaBertozzi02}.  Note that one of the features 
characterizing  
depletion can be the degree of velocity-vorticity alignment or 
"Beltramization" of the flow.

Up to now singularities in the complex domain were studied in Eulerian 
coordinates. In the present paper we analyse some aspects related to 
singular behaviour of flows extended into
the complex domain in Lagrangian coordinates.
One practical motivation of studying Lagrangian singularities 
in the complex domain concerns the structure of Eulerian singularities
of a passive scalar.
Indeed, consider a passive scalar\footnote{Similar remarks can be made  
about a
passive magnetic field.} advected by an incompressible flow $\v$, whose
density satisfies the continuity equation
\begin{equation} 
\partial _{t} \theta (\x,t) + \nabla\cdot (\theta\v) = 0,
\label{scalpass}
\end{equation}
as long as we can neglect molecular diffusion. 
Obviously, this can be solved by using Lagrangian coordinates:
\begin{equation}
\theta(\x,t) = \theta_0(\a(\x,t)),
\label{thetasol}
\end{equation}
where $\theta_0$ is the initial density field and  $\a(\x,t)$ is the inverse
of the Lagrangian map $\x(\a,t)$, the latter satisfying the characteristic 
equation
\begin{equation}
\dot \x =\v(\x,\, t), \qquad \x(\a,\, 0)=\a.
\label{charact}
\end{equation}
If $\theta_0$ is devoid of singularities (entire function), it is clear
that the singularities of the Eulerian field, continued to complex 
locations,
$\theta(\z,t)$, will correspond to those (complex) fluid particles 
trajectories
which have started at $t=0$ at (complex) infinity.
Furthermore, it implies that the width of 
the analyticity strip $\delta (t) $ 
for $\theta (\x ,t)$ in Eulerian coordinates
equals the width of the
analyticity strip $\delta _{\L } (t) $ of the Lagrangian map $\x(\a,t)$
for the time reversed flow. 
In Section~\ref{s:shortlong} we will use this correspondence 
to define 
the smallest scale of the passive scalar field $\theta (\x , t) $ at 
time $t$.

Note that the above argument can be applied to the 2-D vorticity. 
Assuming infiniteness of Eulerian vorticity at complex Eulerian 
singularities
\citep{FMB03, MBF03} it implies that the Eulerian singularities of solutions
of the 2-D Euler equation starting from entire initial data come from 
the complex infinity.
Therefore, especially for the 2-D flows, some understanding of complex 
singularities in Lagrangian coordinates, even for a very simple flow,
can shed light on the nature of Eulerian complex singularities.

S.~\citet{orszag2003} pointed out that the phenomenon of depletion is 
intrinsically Eulerian and has no Lagrangian counterpart\footnote{There are
indications that depletion in 
the 3-D Lagrangian coordinates exists \citep{Ohkitani02},
nevertheless it seems
to be weaker than in the Eulerian coordinates. Furthermore,
our definition of depletion \citep{Frisch95} is somewhat 
different from Ohkitani's definition of depletion, measuring the preference
of the vorticity for being aligned with a certain eigendirection of
the rate-of-strain tensor (the Eulerian case) and the Cauchy-Green tensor
(the Lagrangian case).}. Thus, it is expected
that Lagrangian singularities are stronger - or at least closer to
the real domain - than Eulerian ones. An extreme form
of depletion\footnote{Observe that the nonlinear terms of the 2-D
and 3-D Euler equations completely vanish for the flows (\ref{psicellular})
and the ABC flow, respectively.} is to work with
a Beltrami flow given by
a steady solution of the incompressible
Euler equation such as  the ABC flow in three dimensions or 
the two-dimensional cellular flow with stream function
\begin{equation}
\psi (x_{1} , x_{2} ) =\sin x_{1} \cos x_{2}
\label{psicellular}
\end{equation}
and velocity
\begin{equation}
v_1= -\sin x_{1} \sin x_2, \quad v_2= -\cos x_{1} \cos x_2.
\label{vcellular}
\end{equation}
The latter is much simpler, because the Lagrangian map can be obtained
explicitly in terms of elliptic functions \citep[Appendix A]{DFGHMS}.  
Moreover, it represents a special case ($A=1$, $B=1$, and $C=0$) of the 
ABC flow, obtained by applying a transformation of coordinates of the type 
used in \citet[Appendix A]{DFGHMS} and reducing the nontrivial dynamics
to two dimensions.
Our goal in this paper will be to obtain for this flow the complex
singularities in Lagrangian coordinates.  

The Lagrangian velocity\footnote{We
cannot work with the Lagrangian vorticity since it remains constant along
trajectories and thus constant in Lagrangian coordinates; the same applies to
the stream function for any steady two-dimensional flow.} is defined
just by the change of variables, that is, as $\v_{_{\rm L}}(\a,t) \equiv
\v(\x(\a,t))$ where $\x(\a,t)$ is the solution of (\ref{charact}) with the
velocity given by (\ref{vcellular}).  Of course, such flow being steady,
completely lacks Eulerian singularities (except at complex infinity).
However, as we shall see, it develops {\it real} Lagrangian singularities in
infinite time.

The outline of the paper is as follows. In Section~\ref{s:ana}, we explicitly
construct the singular set, the explicit construction of the Lagrangian
map being relegated to Appendix~A. In  Section~\ref{s:numerical} the 
Lagrangian velocity, its Fourier transform and $\delta(t)$ are determined
numerically. Section~\ref{s:shortlong}  is devoted to the short-time and 
long-time asymptotics.
\section{Analytic structure of the singular manifold}
\label{s:ana}

\begin{figure}[ht]
\iffigs
\centerline{
\includegraphics[scale=0.5]{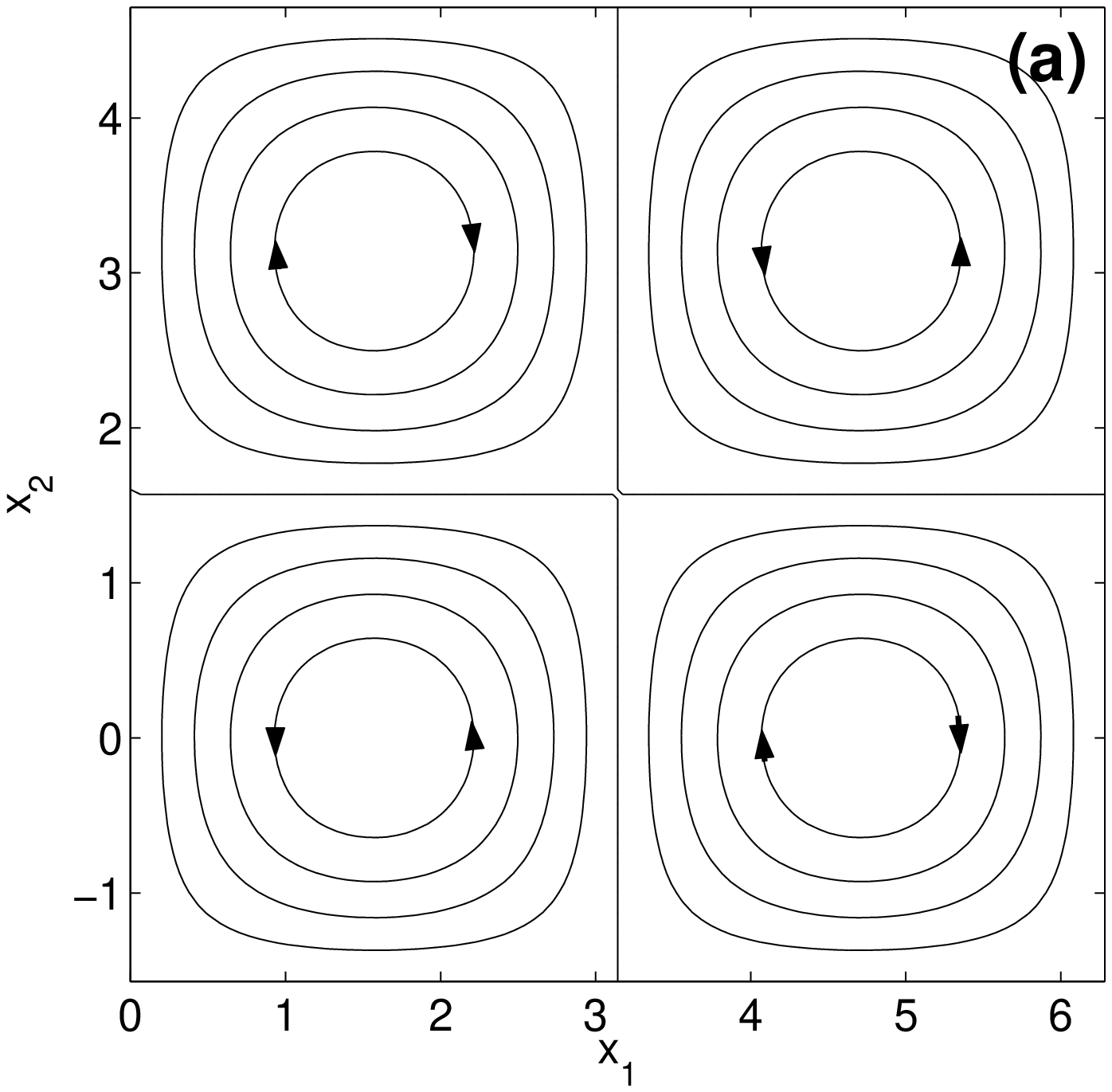}
\includegraphics[scale=0.5]{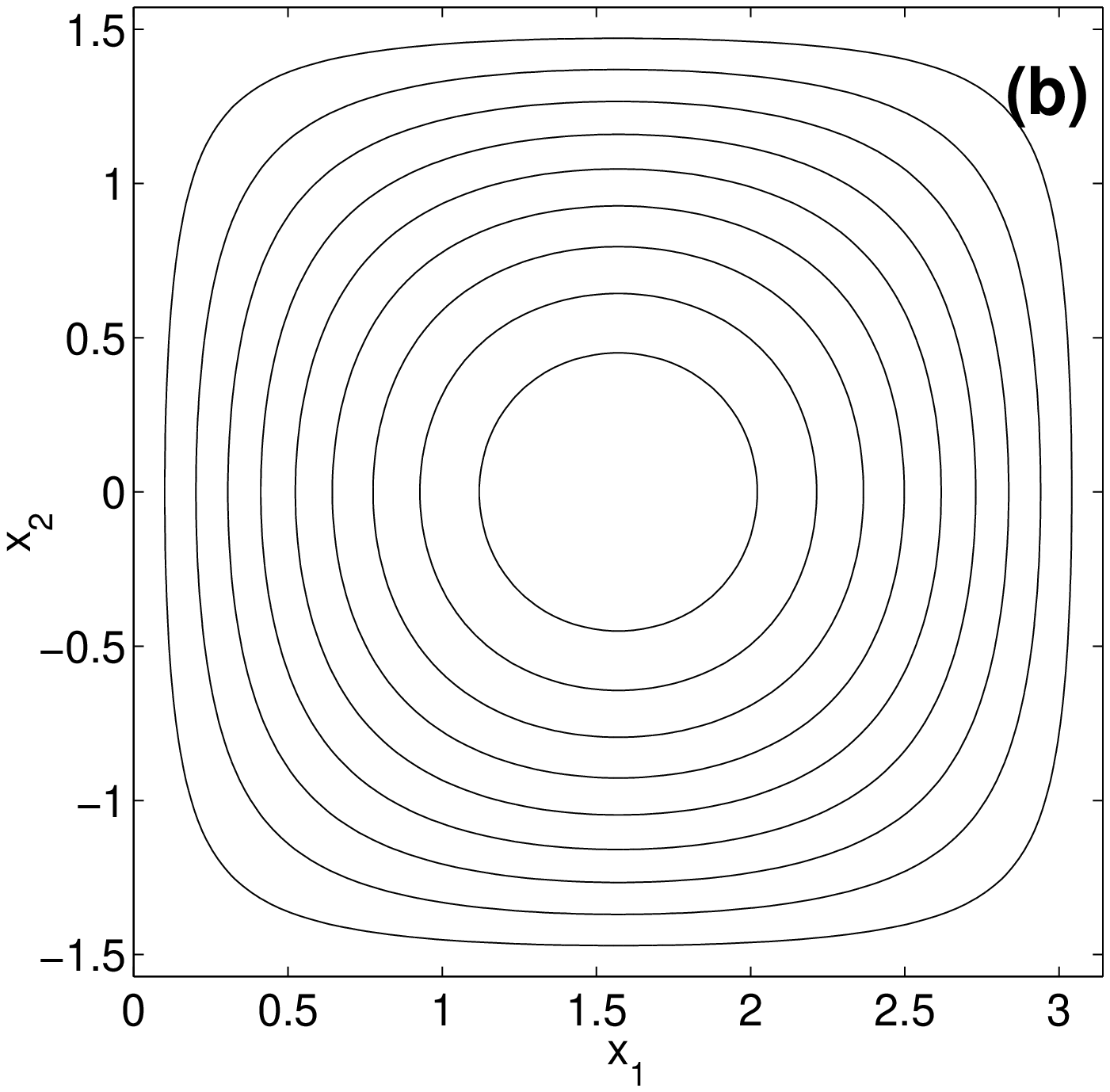}
}
\else\drawing 60 10 {the cellular flow}
\fi
\caption[]{Streamlines for the flow defined 
by (\ref{psicellular}) in the (a) flow box, (b) in the basic cell.}
\label{f:cell}
\end{figure}

We consider the cellular flow on $ \mathbb{T} ^{2} \equiv [0,
2\pi ] \times [0, 2\pi] $\@ defined by (\ref{psicellular}) and
(\ref{vcellular}).
Because of the symmetries of this flow the full periodicity domain
can be decomposed into four cells, Fig.~\ref{f:cell} (a).
For convenience, in Fig.~\ref{f:cell} (b),
we represent the cell $[0,\pi]\times[-\pi/2,\pi/2]$. 
The Lagrangian map $\x(\a,t)$ is the solution of 
\begin{equation} \label{eq 3}
\left\{ \begin{array}{ll}  \dot{x} _{1} = - \sin \ x_{1} \sin \ x_{2}
;\\  
\dot{x} _{2}  = - \cos\ x_{1} \cos\ x_{2} ,
\end{array}
\right.
\end{equation}
with the initial conditions 
\begin{equation} \label{eq 4}
\left\{ \begin{array}{ll}  
x_{1} (\a,\, 0)  = a_{1} ;\\ 
x_{2} (\a,\, 0)  = a_{2} .
\end{array}
\right.
\end{equation}
When fluid particles start at complex locations they will stay in the complex
domain. We then denote the complexified Lagrangian position by $\c=\a+i\b$ and
the complexified Eulerian position by $\z=\x+i\y$. The complex Lagrangian
stream function is $\xi(\c,t) \equiv \psi(\z(\c,t))$ and the complex
Lagrangian velocity is $\v_{_{\rm L}}(\c,t) \equiv \v(\z(\c,t))$.

Since the Eulerian velocity $\v$ is an entire function of $\z$~,\footnote{An
entire function is analytic for all $\z$ but can still be singular at
infinity.} the Lagrangian velocity can become singular only where the
Lagrangian map becomes singular; furthermore it is easily shown that the only
way $\z(\c,t)$ can become singular, for a finite real $t$ and a complex $\c$,
is by going to (complex) infinity.

The explicit solution to (\ref{eq 3}) is given in the Appendix in terms
of elliptic functions. Its form implies the existence of singularities,
associated to certain poles of suitable elliptic functions. This is however
not the simplest way to actually construct the singular set $\Sigma (t)$,
a time-dependent complex manifold of complex dimension one, that is which can 
be parametrized in terms of one complex parameter. For the construction 
it is simpler to observe (i) that the value of the complex stream function 
$\xi$ does not change along fluid particle trajectories and (ii) that
the stream function, being the product of $w_1 \equiv \sin z_1$ and $w_2
\equiv \cos z_2$,
the only way $\z$ can run to infinity while conserving the stream function
is to have either $w_1 \to \infty$ and  $w_2 \to 0$ or conversely.

In terms of the $w_1$ and $w_2$ variables, the equations for 
fluid particle trajectories can be rewritten as
\begin{equation} \label{eq 5}
\left\{ \begin{array}{ll}  
\dot{w} _{1} = - w_{1} \sqrt{1-w_{1}^{2} }\sqrt{1-w_{2}^{2} } \\ 
\dot{w} _{2} = w_{2} \sqrt{1-w_{2}^{2} } \sqrt{1 - w_{1}^{2} } 
\end{array}
\right.
\end{equation}
with now complex initial conditions
\begin{equation}
w_{1} (0) = \sin c_{1} , \qquad  w_{2} (0) = \cos c_{2},
\label{wcrel}
\end{equation}
corresponding to the initial conditions (\ref{eq 4}). 

Obviously, the product $\xi =w_1w_2$ is an integral of motion: this expresses
just that for steady flow particles move along streamlines. One can take 
advantage of this integral of motion to rewrite (\ref{eq 5}) as a single
equation in terms for example of the variable $u\equiv w_1/\xi$, as
\begin{equation} 
\label{eq 9}
\dot{u} = - i\sqrt{1 - u^{2} } \sqrt{1 - \xi ^{2} u^2 }, 
\qquad u(0)=u_{0} = (1/\xi ) \sin c_{1}.
\end{equation}
This equation allows to express the time variable $t$ in terms of the initial
and current values of $u$ by an elliptical integral:
\begin{equation} \label{eq 10}
it = - \int_{u_{0} }^{ u (t) } \frac{ds}{\sqrt{(1-s^{2} )(1-\xi
^{2} s^{2} ) } } .  
\end{equation}

We turn now to the determination of the singularities, that is those initial
complex locations $\c^\star=(c_1^\star,c_2^\star)$ which go to infinity
at time $t$.
Because of the symmetries of the flow we can, without loss of generality,
assume that when $\z\to\infty$ at time $t$, we have $w_{2}
\to  \infty $ and  $w_{1} \to 0$\@ and hence $u(t)=0$. It follows then from 
(\ref{eq 9}) that the singular values of $u_0$ are such that
\begin{equation} \label{eq 11}
i t =-   \int_{u_{0} }^{0} \frac{ds}{\sqrt{(1-s^{2} )(1-\xi
    ^{2}     s^{2} ) } } .
\end{equation}
The inverse of the elliptical integral is the  elliptic 
${\rm sn}$ function \citep{AS65}. Thus $u_0$ can be expressed 
as 
an elliptic function. Since, by (\ref{eq 9}), $w_1(0)=\xi u_0$, we obtain
\begin{equation} \label{eq 12}
w_{1} (0) = \xi \,\mathrm{sn} (it ,\xi ) ,\qquad  w_{2} (0) =
\mathrm{ns} (it ,\xi ),
\end{equation}
where  ${\rm ns} \equiv 1/{\rm sn}$.
Alternatively, since the ${\rm sn}$ and ${\rm ns}$ functions of 
imaginary arguments can be expressed in terms of the ${\rm sc}$ function,
defined as $\mathrm{sc} \equiv \mathrm{sn} / \mathrm{cn} $, and
its reciprocal ${\rm cs}$, we have
\begin{equation} \label{eq 13}
w_{1} (0)= i \xi \, \mathrm{sc} (t ,\sqrt{1 - \xi ^{2} } ) 
, \qquad  w_{2} (0) = - i\,
  \mathrm{cs} (t
  ,\sqrt{1 - \xi ^{2} } ) .
\end{equation}
By (\ref{wcrel}), we can express the singular Lagrangian locations
at time $t$ as
\begin{equation} 
c^{\star }_{1}  =  \arcsin \left[ i\xi \mathrm{sc} 
(t ,\sqrt{1-\xi ^{2} }
  )  
\right] ,\qquad 
c^{\star }_{2}  =  \arccos \left[ i\mathrm{cs} 
(t,\sqrt{1-\xi ^{2} } )
\right] . 
 \label{eq 14}
\end{equation} 
This constitutes the parametric representation of the singular manifold
$\Sigma(t)$,
the parameter being the complex stream function $\xi$. Obviously, the functions
appearing in this representation are multivalued. More precisely, the singular
manifold has infinitely many sheets. 

It is well known that elliptic functions are real for real values of the 
parameters. Therefore, $\Sigma(t)$ will have a non-trivial part
``above'' the 
hyperbolic stagnation points, namely the four corners in Fig.~\ref{f:cell}(b).
\footnote{By ``above'' we mean  at complex locations whose real parts are
the stagnation points.}  By symmetry,
it  is enough to consider one of them. We take $(a_1,a_2)= (0,\pi/2)$
and easily obtain the following parametric representation of the singular
locations having this stagnation point as real part:

\begin{equation} 
\label{eq 15}
b^{\star }_{1} = \mathrm{arcsinh} \left[ \xi \mathrm{sc}
  (t ,\sqrt{1-\xi ^{2} } )  
\right] , \qquad
b^{\star }_{2} = \mathrm{arcsinh} \left[ -\mathrm{cs} 
(t ,\sqrt{1-\xi ^{2} } )  
\right] . 
\end{equation}

\begin{figure}[ht]
\iffigs
\centerline{
\includegraphics[scale=0.42]{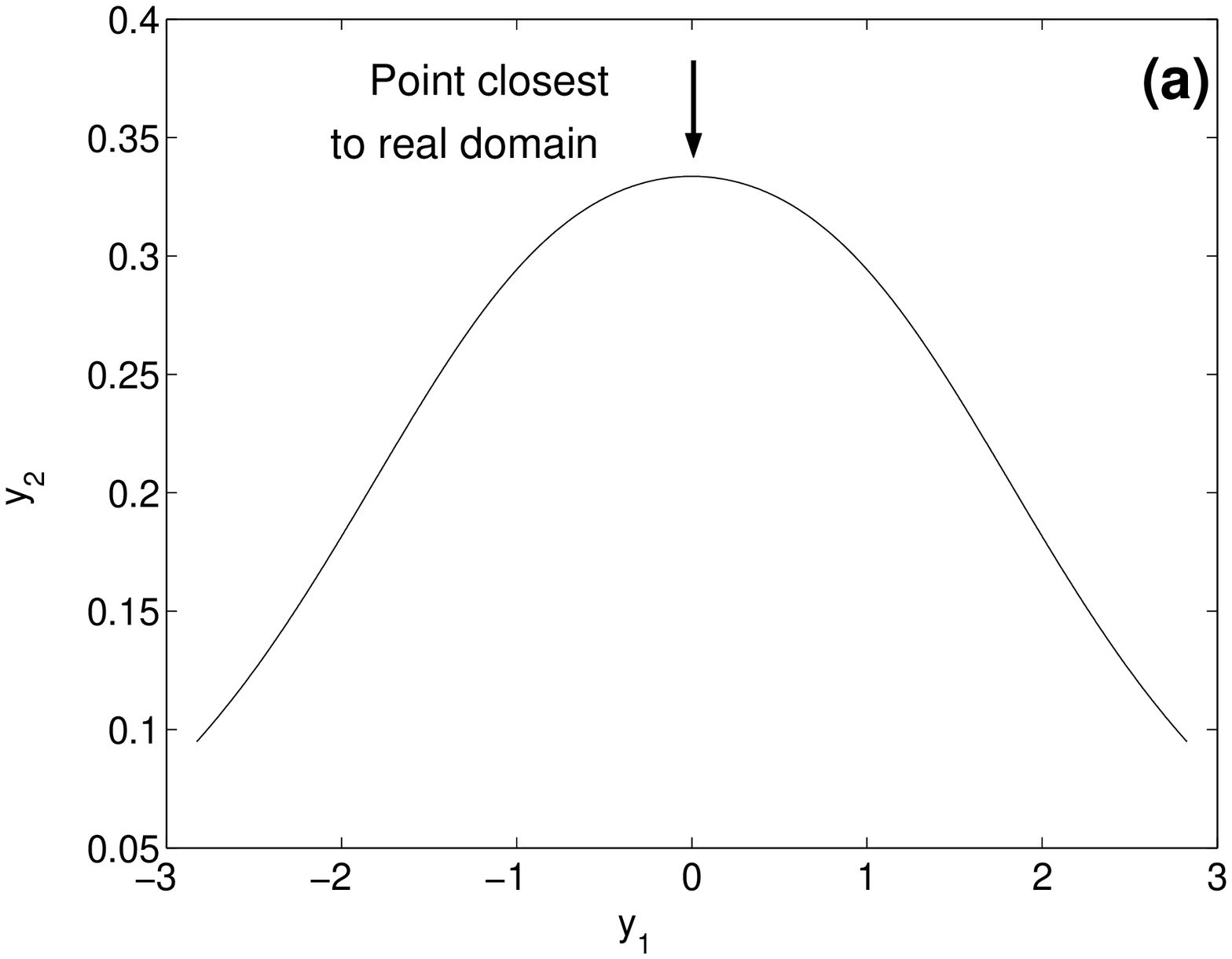}
\includegraphics[scale=0.42]{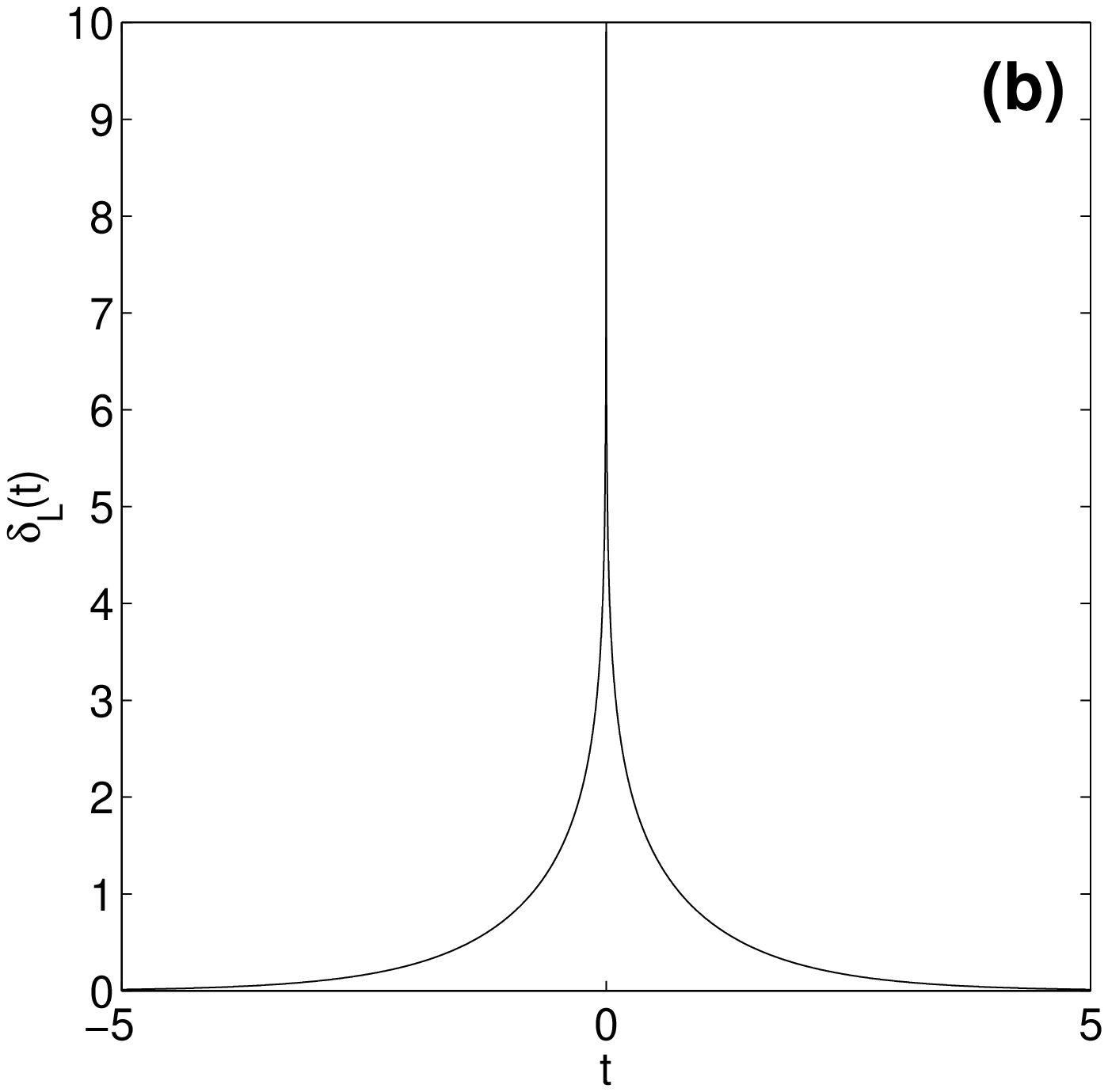}
}
\else\drawing 60 10 {the singular manifold}
\fi
\caption[]{(a) The cut of the singular 
manifold on the imaginary $(y_{1} ,y_{2} ) $-plane is represented\@. 
The point closest to the origin lies on the $y_2$-axis. (b) 
Here the value of the closest distance from singular manifold
to the real domain, $\delta (t) $, is represented for both positive and
negative times.}  
\label{f: trace}
\end{figure}

Using the parametrization (\ref{eq 15})  
in Fig.~\ref{f: trace} (a) we have represented 
the curve obtained 
as  intersection of the singular manifold with the set of points with real
coordinates fixed at $(x_{1} , x_{2} )=(0,\pi /2 )$
, that is the trace of the singular manifold on the pure imaginary
plane lying above the corresponding stagnation point.
Here "trace on" is used with its mathematical meaning of "intersection with".

The width of the Lagrangian analyticity strip $\delta_\L(t)$ is obtained by 
finding
the point(s) on the singular manifold closest to the real domain.
It seems likely that these points will be located over the points 
where the flow
exhibits a special structure. 
Indeed, expanding (\ref{eq 14}) locally
in its Taylor series
confirms  that the points closest to the real domain
lie above the hyperbolic stagnation points. The upper left
and lower right stagnation points have $b_1^\star=0$, while the upper right
and lower left stagnation points have $b_2^\star=0$. The distance from
the real domain is easily shown to be given by
\begin{equation} \label{eq 20}
\delta_\L(t) = \left| \mathrm{arcsinh} 
\left( \frac{1}{\sinh t} \right) \right|.
\end{equation}
Fig.~\ref{f: trace} (b) is a plot of $\delta_\L(t)$ showing both positive and
negative times.

As we have mentioned in the introduction, 
$\delta_\L(t) $ of the time reversed flow
gives the width of the 
analyticity strip in  Eulerian coordinates for a 
passive scalar satisfying (\ref{scalpass}) with the same 
velocity field (\ref{vcellular}). Indeed, singularities for the passive
scalar correspond to (complex) fluid particles which are mapped back to 
infinity in time $t$. 
Note that the symmetry of the cellular  flow used
here implies that $\delta_\L (t)$ is an even function of time so that 
$\delta_\L(t) $ of the original flow coincides with the Eulerian 
$\delta (t) $ for the passive scalar field.

\section{Numerical integration in Lagrangian coordinates}
\label{s:numerical}

In this section we obtain the  width $\delta_\L(t)$ of the Lagrangian
analyticity strip by numerically calculating the Fourier transform of 
the velocity in Lagrangian coordinates and then applying the
method of tracing complex singularities \citep{SSU83} to relate the Fourier
transform to $\delta_\L(t)$. Our method is  implemented for the
steady cellular flow given by (\ref{psicellular}) but can in 
principle be applied to any steady solution of the Euler equation in both two
and three dimensions. 

We need to calculate the Lagrangian Fourier coefficients, which
are here obtained as follows.
First, we calculate the Lagrangian map $\x(\a, t)$
by solving the characteristic equations (\ref{eq 3})
with a fourth order Runge-Kutta
method for $N^2$ initial conditions
$\a = ((2\pi/N) l,\, (2\pi/N) m)\,\,(l,\,m = 0,1,\ldots, N - 1)$,
which form a regular square grid in the Lagrangian marker space.
We obtain then the Lagrangian velocity field by just changing variables:
\begin{eqnarray}
\v_\L(\a, t) \equiv  \v(\x(\a,\ t),\ t),
\label{e:lagvel} 
\end{eqnarray}
where $\v(\x)$ is given by (\ref{vcellular}).
The Lagrangian velocity Fourier coefficients are given by
\begin{eqnarray}
 \hat{\v}_\L(\k,\ t) 
= \frac{1}{N^2}\sum_{\a} \v_\L(\a, t)\, e^{-i\k\cdot\a}.
\label{e:lagvelF} 
\end{eqnarray}
For measuring the width $\delta_\L$, it is convenient to define
the shell-summed amplitude $A^{(\v_\L)}_k$ of  $\hat{\v}_\L(\k)$ as 
\begin{equation}
 A^{(\v_\L)}_k = \sum_{k \le |\k| < k + 1} |\hat{\v}_{\L}(\k)|.
\end{equation}
The Lagrangian width $\delta_\L$ is then obtained by fitting the
shell-summed amplitude to an exponential with an algebraic prefactor:
\begin{equation}
  A^{(\v_\L)}_k \propto k^{-\alpha}\exp(-\delta_\L k).
\label{e:decrement}   
\end{equation}

\begin{figure}
\iffigs
\includegraphics[scale = 0.75]{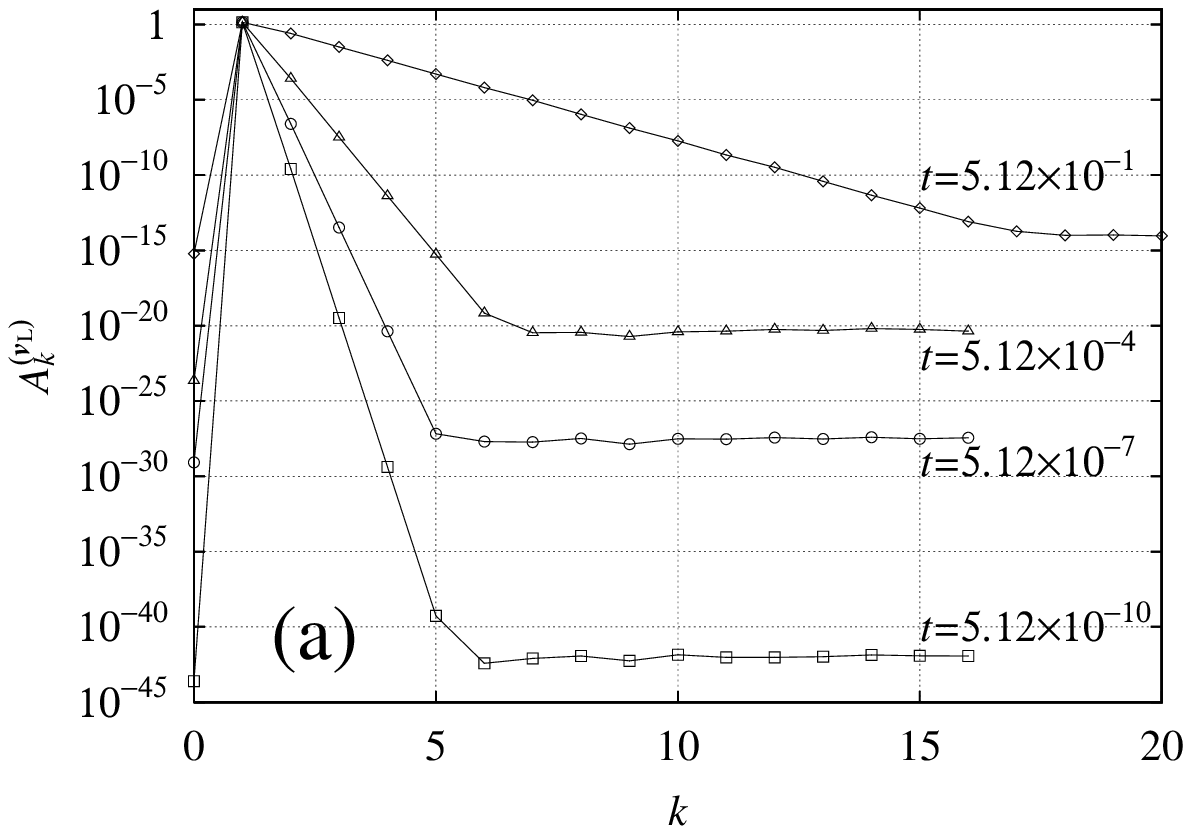}\\
\includegraphics[scale = 0.75]{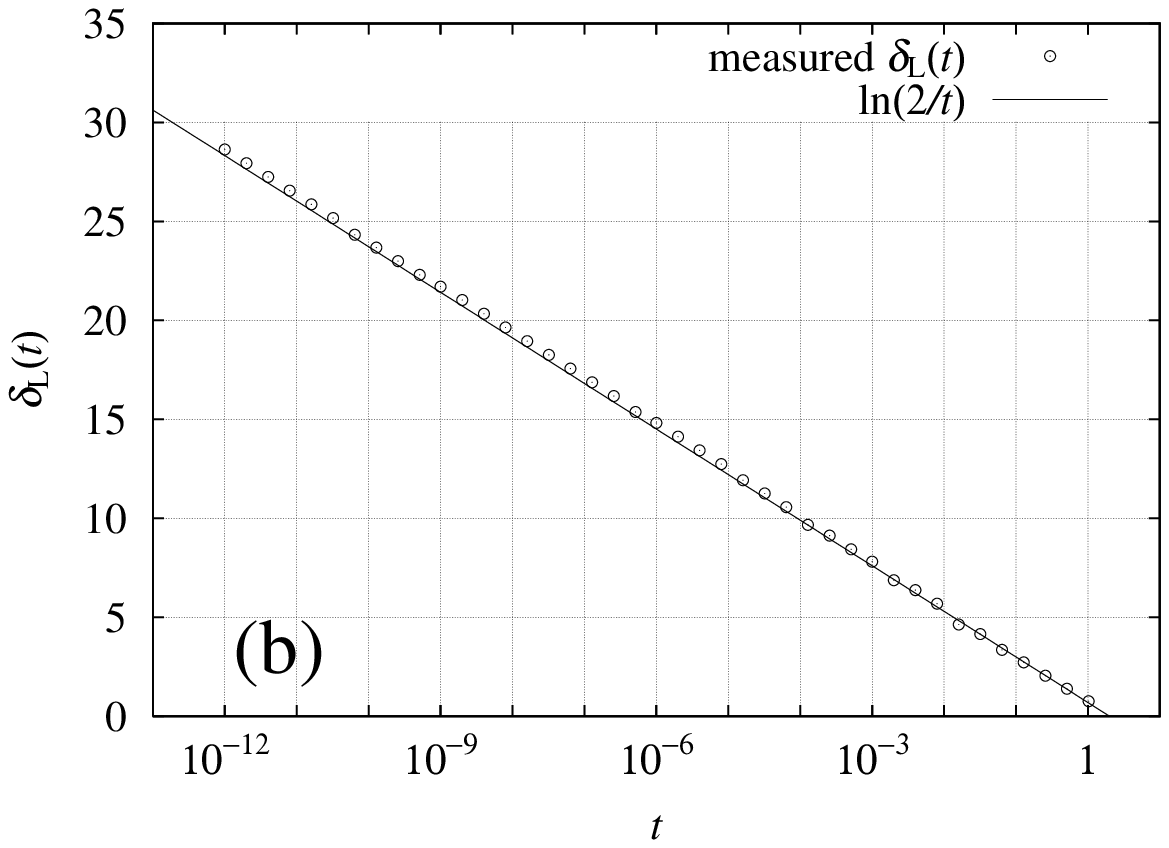}
\else\drawing 65 10 {short time ssa and delta}
\fi 
\caption{\label{f:short_ssa_delta}
 (a) Shell-summed amplitudes of the Lagrangian velocity $A^{(\v_\L)}_k$ for
 four different short times calculated with the number of grid points $32^2$ and $64^2$.
 (b) Log-linear plot of the width of the Lagrangian analyticity strip $\delta_\L(t)$ measured
 as the logarithmic decrement of $A^{(\v_\L)}_k$ at short times $10^{-12} \le t \le 1$.
}
\end{figure}

We first discuss how to obtain the width $\delta_\L(t)$ at short times when
it is very large (since it is infinite at $t=0$). Obtaining  $\delta_\L(t)$
with a double-precision calculation is not feasible because the Fourier
amplitudes $A^{(\v_\L)}_k$ fall off too quickly as a function
of the wavenumber and thus get lost in the roundoff noise.
So we need to solve the equation (\ref{eq 3}) with higher precision.
\footnote{We use here the package MPFUN90 \citep{B95}.}
In practice we divide the time interval $10^{-12} \le t \le t$ into four parts:
(i)  $10^{-12} \le t < 10^{-9}$,
(ii) $10^{-9}  \le t < 10^{-6}$,
(iii)$10^{-6}  \le t < 10^{-3}$,
(iv) $10^{-3}  \le t < 1$.
For each part, we use different time step $\Delta t$ and precision:
(i)  $\Delta t = 10^{-12}$ and 48-digit precision,
(ii) $\Delta t = 10^{-9}$  and 36-digit precision,
(iii) $\Delta t = 10^{-6}$  and 24-digit precision,
(iv)  $\Delta t = 10^{-3}$  and 15-digit (double) precision.
These precisions, for each $\Delta t$, are the best we can handle with
the fourth-order Runge-Kutta scheme.
Note that in \citet{FMB03}   a 90-digit spectral calculation was used
for obtaining the short-time behavior of $\delta(t)$ in Eulerian
coordinates for  2-D Euler flow with non-trivial Eulerian dynamics; in the
present case, the accuracy of the shell-summed amplitudes $A^{(\v_\L)}_k$
is constrained by the accuracy of the time integration scheme of 
(\ref{eq 3}).
\footnote{%
We cannot take higher precision than the order of $(\Delta t)^4$, which is the error
level of the fourth-order Runge-Kutta method employed here.
If we take higher precision than that, we find that
a bump appears in the shell-summed amplitude $A^{(\v_\L)}_k$
around the level of $(\Delta t)^4$.
This is perhaps due to the fact that the equation (\ref{eq 3}) is integrated
in the physical space. In \citet{FMB03}, the Euler equation was integrated
with the same fourth order Runge-Kutta scheme but in the Fourier space,
and this may considerably decrease the error (more precisely, the factor in front
of $(\Delta t)^4$ is exponentially small for large $k$).
}%
 Also we encounter the difficulty that a long-time integration causes
serious accumulation of error in vorticity, which
eventually breaks the conservation of vorticity (in 2-D  
vorticity should be conserved along each fluid particle trajectory).
That is the reason why we split the time interval
$10^{-12} < t < 1$ into four parts and use different $\Delta t$ for each part
to avoid this accumulation.
Figure \ref{f:short_ssa_delta} shows the shell-summed amplitude $A^{(\v_\L)}_k$
and the width of the Lagrangian analyticity strip $\delta_\L(t)$
at short times $10^{-12} \le t \le 1$.
As seen in Fig.~\ref{f:short_ssa_delta} (a),
the number of points in $A^{(\v_\L)}_k$ to be fitted is small.
We checked that, for several measured $\delta_\L(t)$'s shown in Fig.~\ref{f:short_ssa_delta} (b),
a calculation with smaller $\Delta t$ and higher precision
for the same instance gives the same value of $\delta_\L(t)$.
In this sense, we believe
the measured $\delta_\L(t)$ is reliable in spite of the very limited data points.
From Fig.~\ref{f:short_ssa_delta} (b), a clean logarithmic decay of $\delta_\L(t)$
is observed for more than 10 decades.

\begin{figure}
\iffigs
\includegraphics[scale = 0.75]{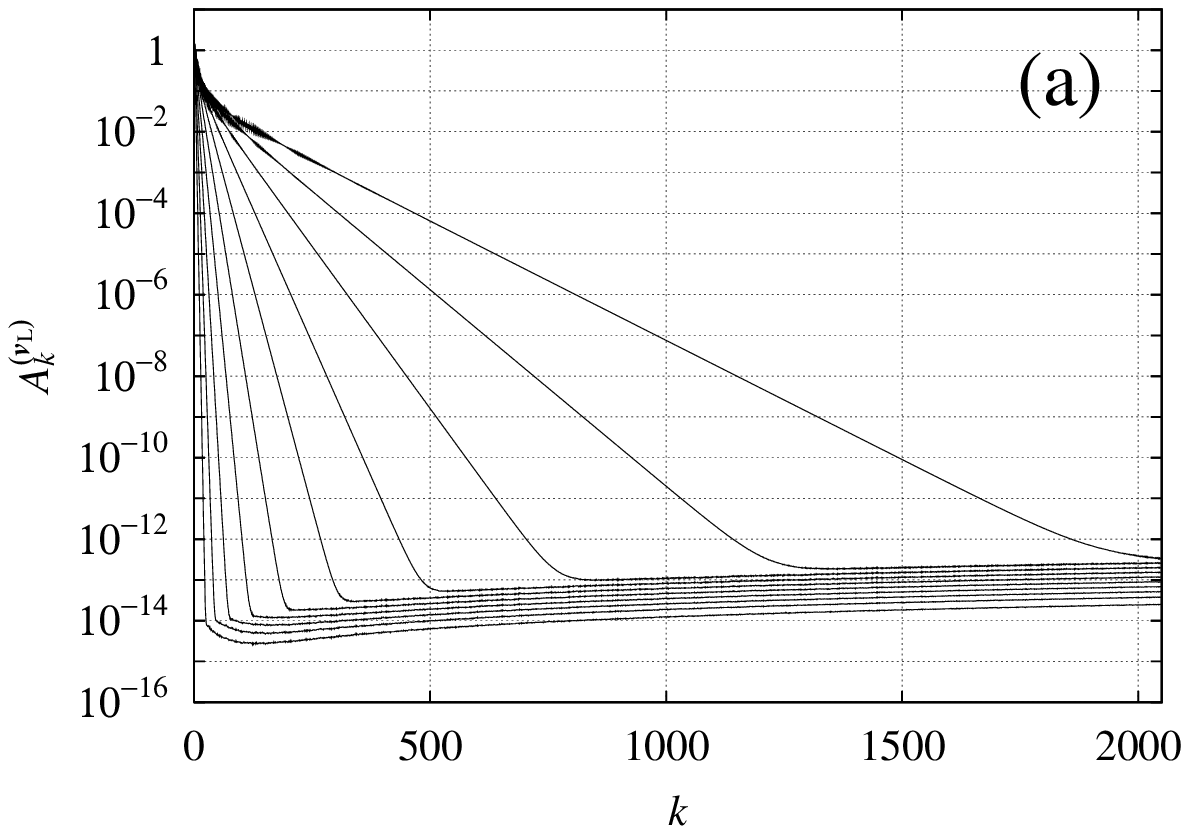}\\
\includegraphics[scale = 0.75]{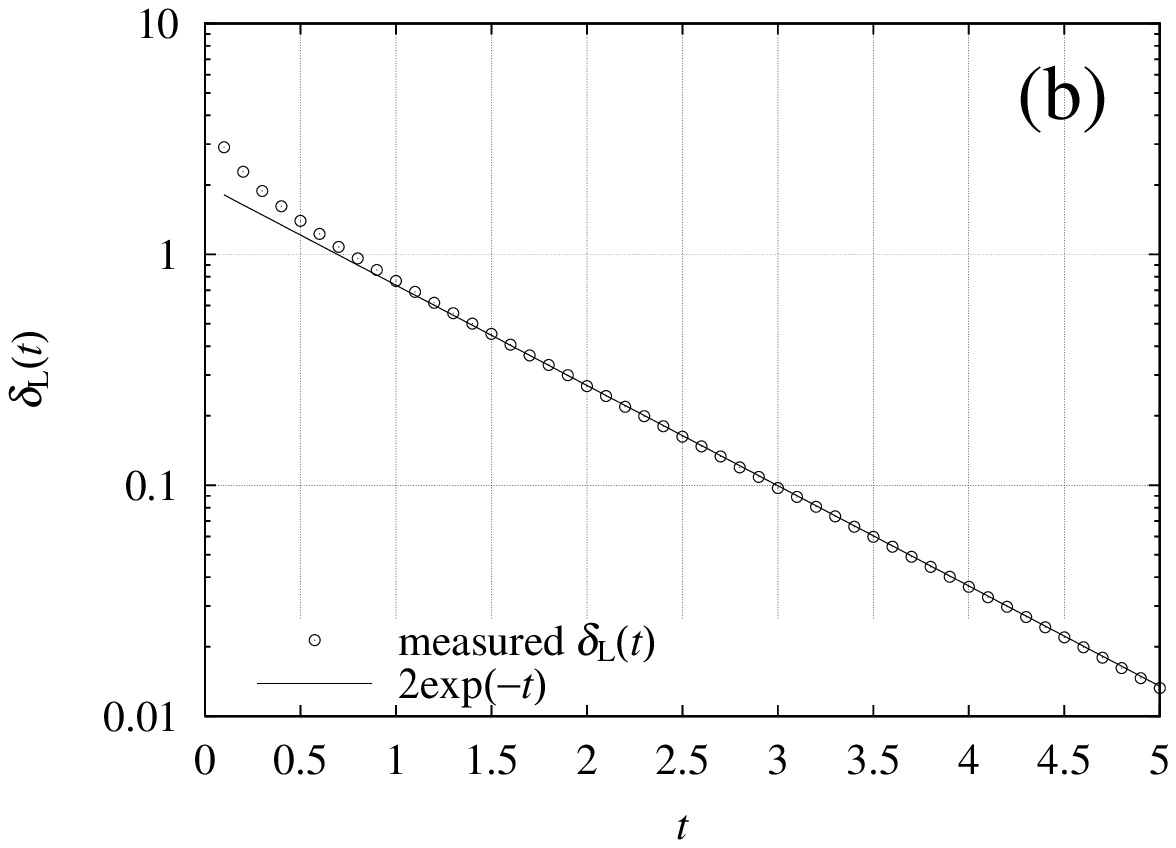} 
\else\drawing 65 10 {long time ssa and delta}
\fi 
\caption{\label{f:long_ssa_delta}
 (a) Shell-summed amplitudes of the Lagrangian velocity $A^{(\v_\L)}_k$ for
 10 different instances ($t = 0.5, 1.0,\ldots, 5.0$) calculated with the number of grid points $4096^2$.
 (b) Linear-log plot of the width of the Lagrangian analyticity strip $\delta_\L(t)$ measured
 as the logarithmic decrement of $A^{(\v_\L)}_k$ at long times $t \le 5$.
}
\end{figure}

Now we turn  to the numerical determination of the 
long-time behavior of the width of the Lagrangian analyticity
strip. The calculation is straightforward: the characteristic 
equation (\ref{eq 3}) is solved
with standard double precision using  a time step $\Delta t = 10^{-3}$.
The number of grid points is $4096^2$. Figure \ref{f:long_ssa_delta}
shows $A^{(\v_\L)}_k$ and $\delta_\L(t)$ at long times, $0 < t \le 5$.
The temporal decrease of  $\delta_\L(t)$ is exponential in this regime.

\begin{figure}
\iffigs
\centerline{%
\includegraphics[scale = 0.4, angle=-90]{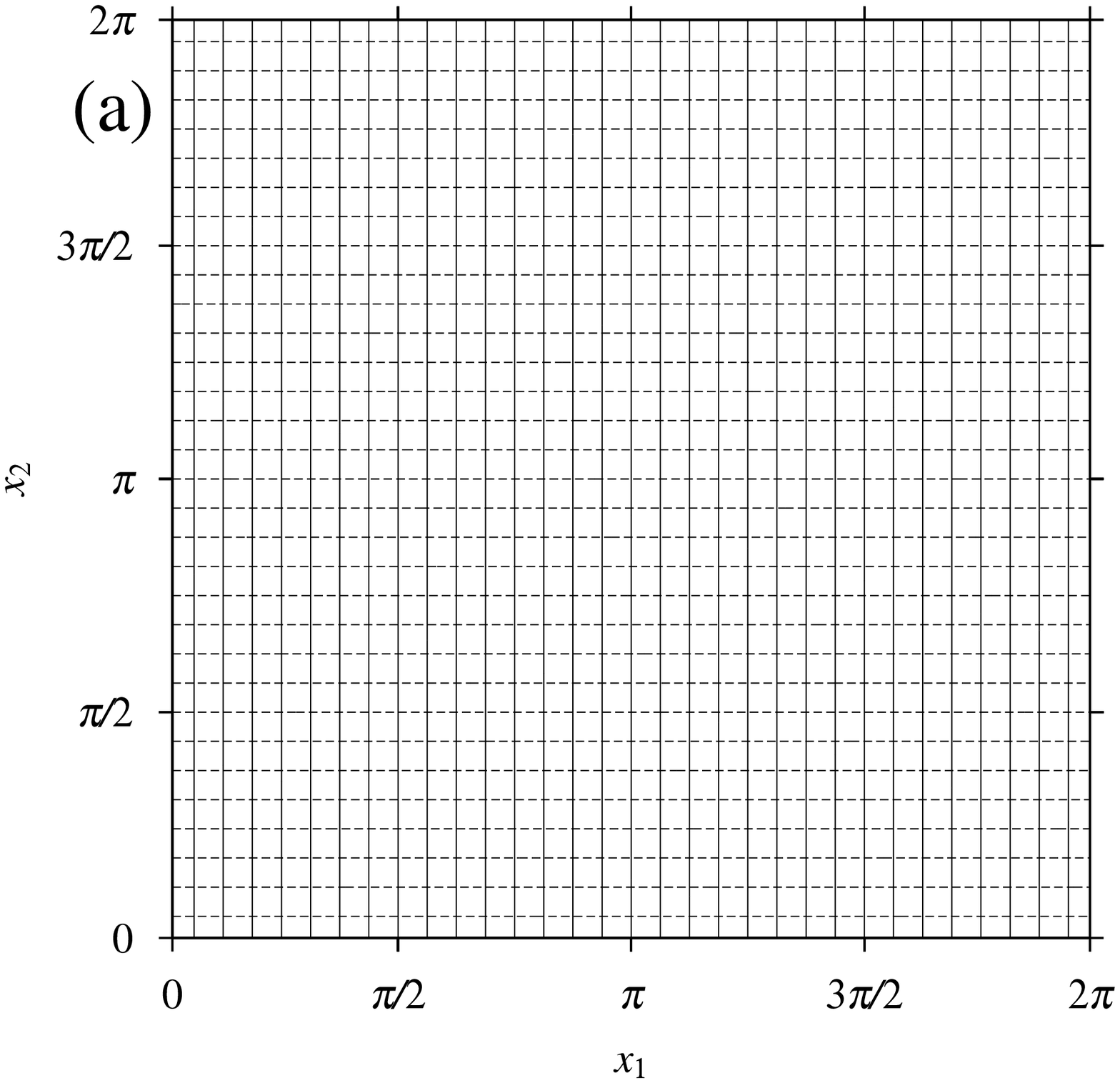}
\includegraphics[scale = 0.4, angle=-90]{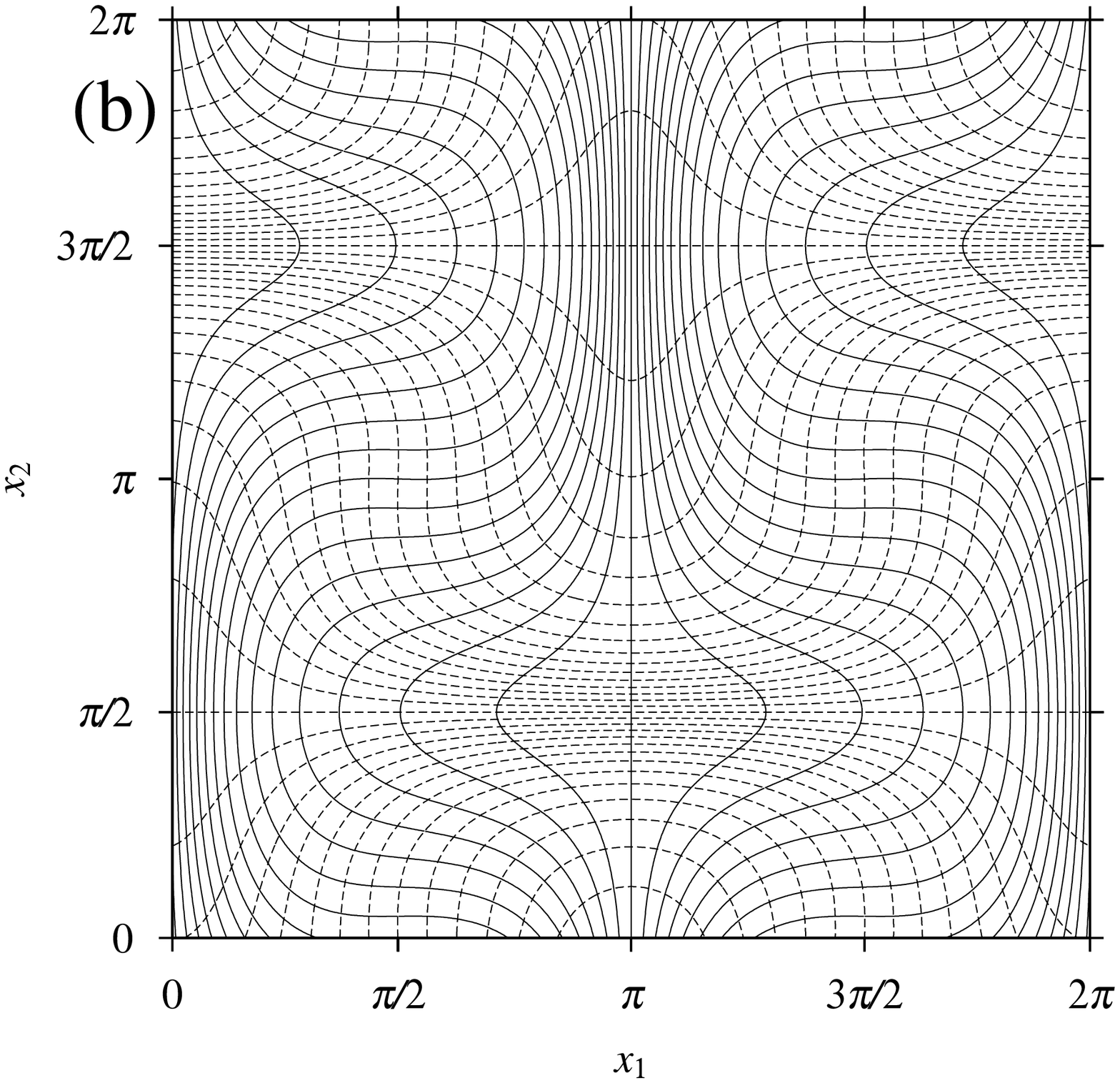}
}

\hspace*{5mm} 
 
\centerline{ 
\includegraphics[scale = 0.4, angle=-90]{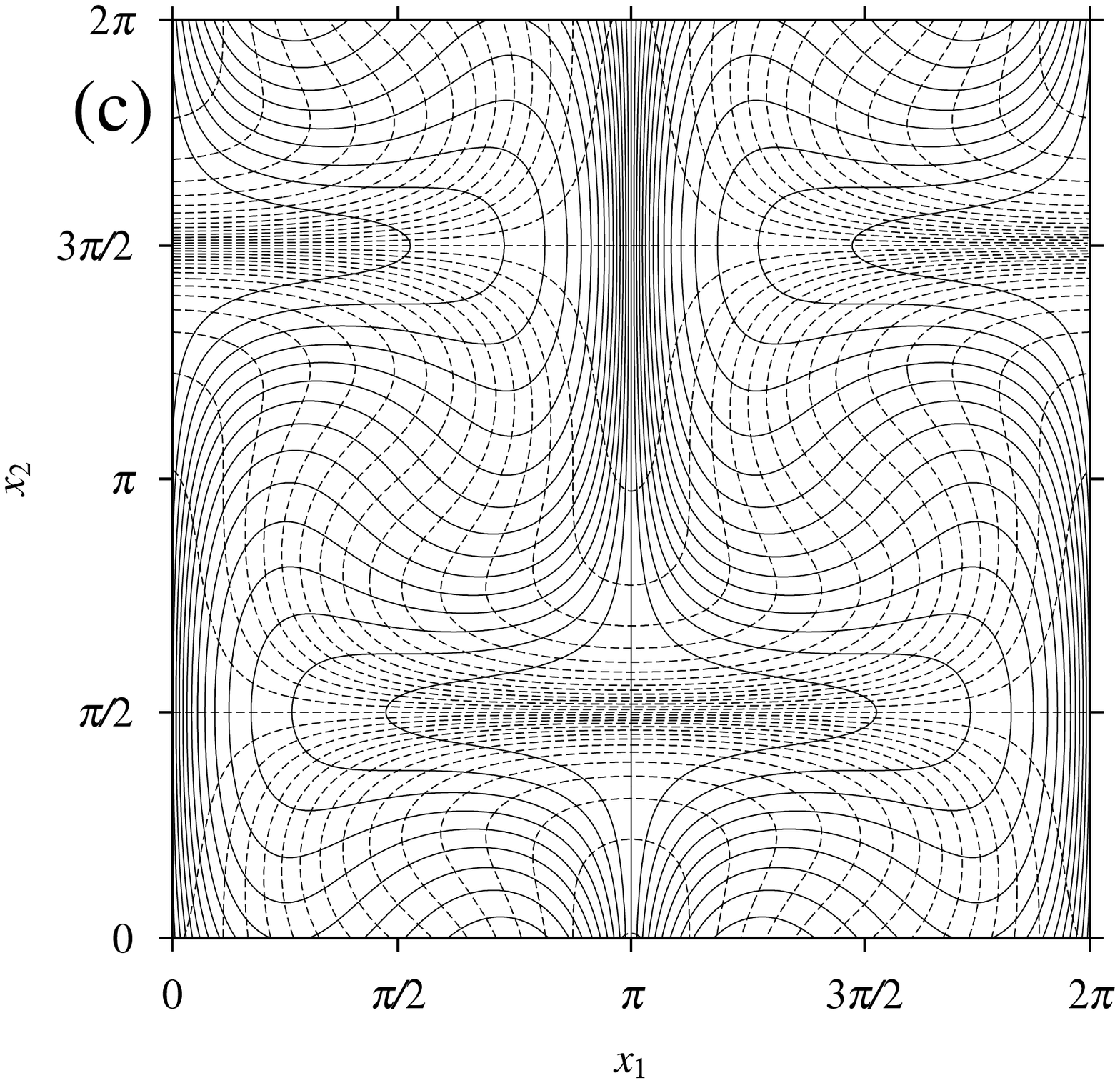}
}%
\else\drawing 65 10 {advected mesh}
\fi 
\caption{\label{f:mesh}
 (a) Regular grids  in the Eulerian coordinates $(x_1,\, x_2)$ at time $t = 0$.
 (b) Advected regular grids by the steady flow at time $t = 1.6$, which
    indicate the Lagrangian map $\x (\a, t)$ in the Eulerian coordinates.
 (c) Advected regular grids at time $t = 2.4$. 
}
\end{figure}

\begin{figure}
\iffigs
\centerline{%
\includegraphics[scale = 0.4, angle=-90]{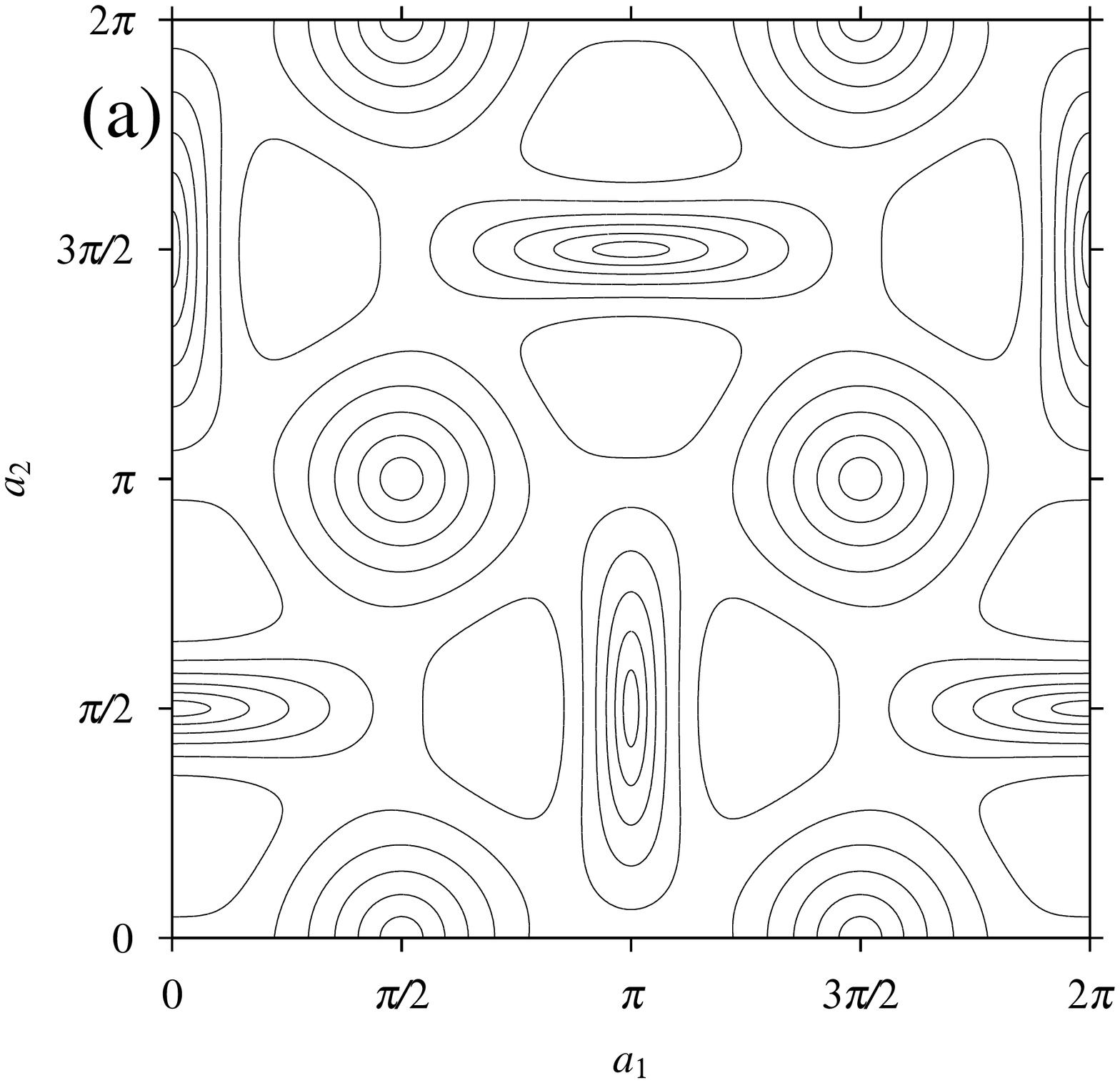}
\includegraphics[scale = 0.4, angle=-90]{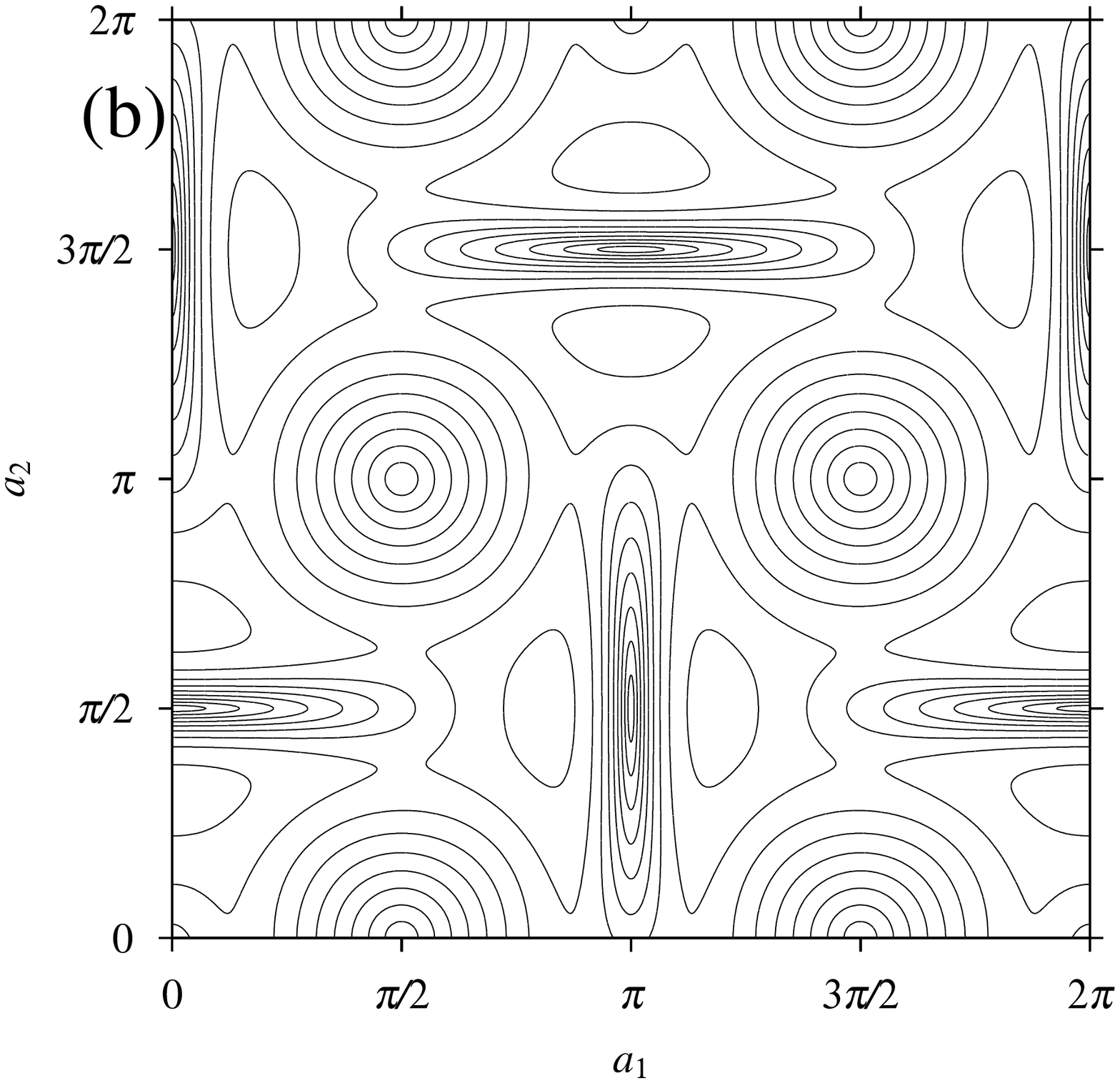}
}
 
\hspace*{5mm}
 
\centerline{ 
\includegraphics[scale = 0.4, angle=-90]{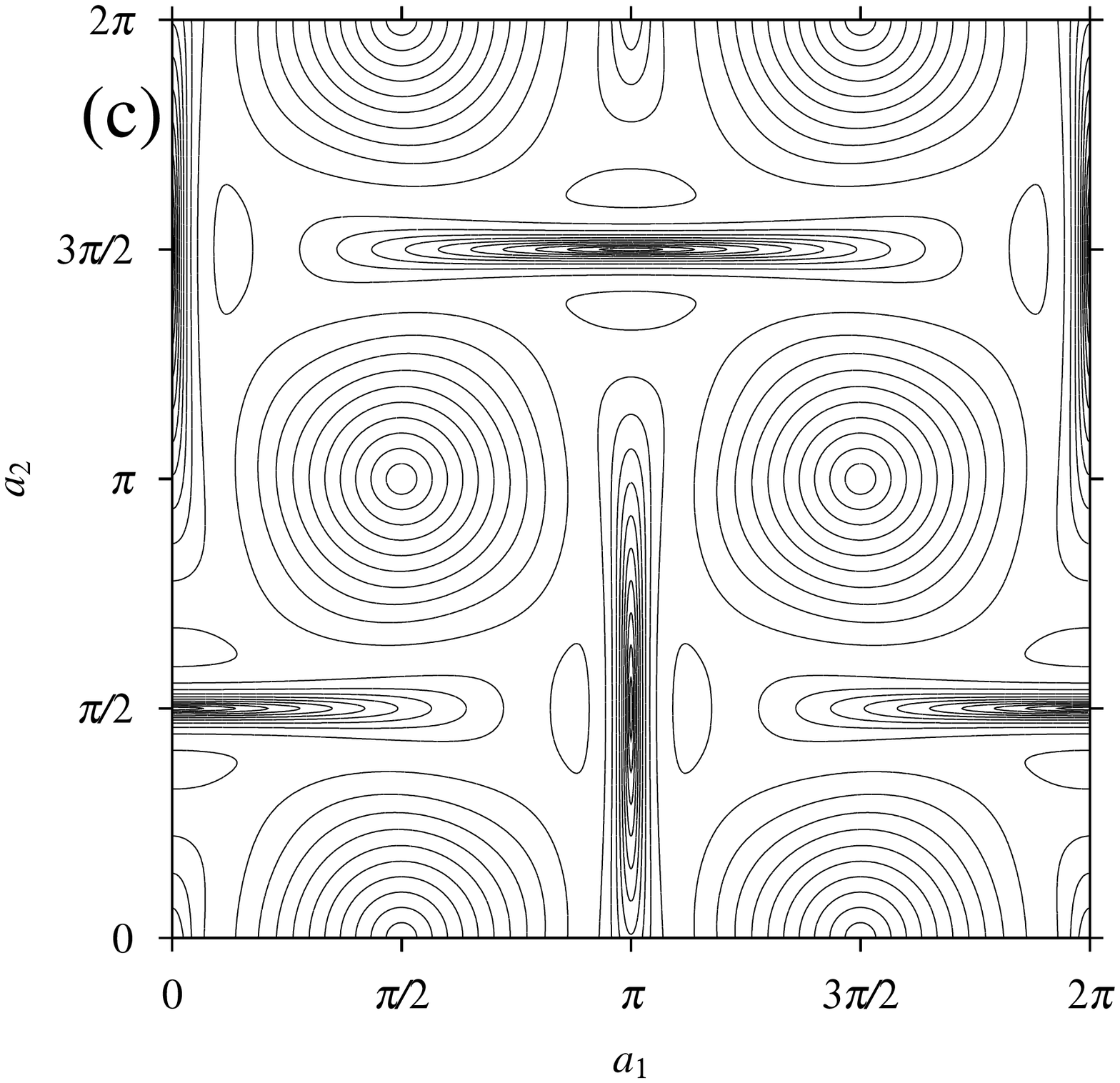}
}%
\else\drawing 65 10 {contour of the displacement}
\fi 
\caption{\label{f:displacement}
 (a) Contours of the modulus of the displacement $|\x(\a, t) - \a|$ 
     at time $t = 1.6$ in the Lagrangian coordinates $(a_1,\, a_2)$.
     Contour levels are $\pi/15,\, 2\pi/15, \ldots$.
 (b) Same as (a) but for $t = 2.4$.
 (c) Same as (a) but for $t = 3.2$.
}
\end{figure}

This indicates that exponentially small-scale structure is generated in
the Lagrangian coordinates by this steady flow.
For seeing this, we first show in Fig.~\ref{f:mesh}
the deformation of the initially regular grid caused by the flow.
As seen in Fig.~\ref{f:mesh}, the Eulerian field is squeezed
around the hyperbolic stagnation points along the contracting
direction.  We then plot contours of the modulus of
the corresponding displacement $|\x(\a, t) - \a|$ in the Lagrangian coordinates
for various instances in Fig.~\ref{f:displacement}.
Here we can clearly see that the small-scale structures centered 
at hyperbolic stagnation points are generated in the Lagrangian
coordinates.
The exponential decrease of $\delta_L(t)$
corresponds to the exponential decrease of the width of
the structure in the Lagrangian coordinates centered at
the hyperbolic stagnation points.

\section{Short-time and long-time behavior}
\label{s:shortlong}

The parametric representation (\ref{eq 14}) for the singular manifold at
arbitrary times is somewhat cumbersome. At short times we  can use suitable
expansions of the trigonometric and elliptic functions \citep{AS65} 
and obtain to leading order
\begin{equation} \label{eq 21}
c_{1}^{\star } \backsimeq \mathrm{arcsin} (i\xi t ) ,\qquad 
c_{2}^{\star } \backsimeq \frac{\pi }{2} + i\,\ln \left( \frac{2}{t} \right) ,
\end{equation}
from which it follows readily that the width of the analyticity strip is
\begin{equation} \label{short-time delta}
\delta_\L (t) \backsimeq \ln \frac{2}{t}, \qquad t\to 0 .
\end{equation}
Hence the Lagrangian $\delta_\L(t)$ follows a logarithmic law at short time.

We now observe that a short-time logarithmic law for the Eulerian
$\delta(t)$ has been obtained by \citet{FMB03} (see also
\citet{MBF03}) for 2-D flow with a finite number of initial Fourier
harmonics.  In the present case the Eulerian dynamics are trivial
since the flow is time-independent. We believe that a logarithmic law
for the short-time behavior of the Lagrangian $\delta_\L(t)$ is quite
general for \textit{steady} solutions of the Euler equation with a
finite number of harmonics in any dimension $d\ge 2$. Roughly, the
argument is as follows. Let $n$ be the degree of the highest-order
Fourier harmonic; the velocity at complex locations a large distance
$y$ from the real domain grows as $e^{ny}$. Singularities in
Lagrangian coordinates at time $t$ correspond to fluid particles which
emanated initially from infinity or, equivalently, which are mapped to
infinity in a time $t$ under the reversed flow. The latter also grows
as $e^{ny}$ at large $y$. A fluid particle located initially within a
distance $\delta_\L \gg 1$ of the real domain and having an imaginary
component of the velocity $O\left(e^{ny}\right)$ will escape to
infinity in a time approximately given by
\begin{equation} \label{escape time}
t=\int_{\delta_\L }^{\infty } e^{-n y} dy = \frac{1}{n } e^{-n
  \delta_\L } .
\end{equation}
It follows that 
\begin{equation} \label{generic delta}
\delta_\L (t) \backsimeq \frac{1}{n } \ln \frac{1}{n t}.
\end{equation}
It is not clear whether this law for the Lagrangian $\delta_\L$ will carry over 
to flow having non-trivial Eulerian dynamics.

We now turn to the long-time behavior. In Eulerian coordinates 
for non-trivial dynamics the best proven lower bound for $\delta(t)$
is a decreasing double exponential (see \citet{FMB03} and references
therein). But numerical simulations usually give a simple exponential
decrease \citep{SSU83,FMB03}. The Lagrangian $\delta_\L$ for the 
steady cellular flow considered here is given by (\ref{eq 20}), which
has an obvious expansion at large $t$
\begin{equation} \label{eq:exp fall-off}
\delta_\L (t)  \backsimeq 2e^{-t}, \qquad t\to \infty ,
\end{equation}
which agrees with the numerical simulations of Section~\ref{s:numerical}.
Note that the nearest singularities are located ``above''  the hyperbolic 
stagnation  points of the flow (\ref{vcellular}).

As we have remarked in Section
\ref{s:intro} $\delta _\L (t) $ allows us to give
an objective definition of the smallest scale for the passive scalar 
advected by the flow (\ref{vcellular}). According to formula 
(\ref{eq:exp fall-off}) this scale decreases exponentially 
in time which, as we have noted in Section \ref{s:numerical}, reflects
the squeezing of the flow around the hyperbolic stagnation points.
The exponential temporal rate of deformation of passive scalar structures
transversal to the streamlines can be explained by dimensional reasoning.

We note that the period of motion of a particle along a trajectory
with the stream function $\Psi (\a ) $ is given by 
$4K(\sqrt{1-\Psi ^2 (\a ) } ) $  
(see Appendix \ref{a:explicit}, formula (\ref{eq:solution}) and 
\citep{Lawden}) changing from $2\pi $ in the centre of the cell to 
infinity at the boundaries. Obviously, for a passive scalar field which
is not constant\footnote{If a passive scalar field is constant along the
  streamlines there will be no mixing.}  
along the streamlines, such as for example 
$\theta (x_1, x_2)=\cos x_1 \sin x_2 $, the most intensive mixing will
be observed near the hyperbolic stagnation points.
Let us consider the flow near the stagnation point 
$(0,\pi /2)$ at the point $\a =(a_1 , \pi /2 - a_2)$. 
The displacement of a fluid particle starting at $\a$
becomes significant at the time $ t \sim  K(\sqrt{1-\Psi ^2 (\a ) } ) $. 
Since $\Psi (\a ) $ is small, we use the approximative expression
(see \citep{WW27})
\begin{equation} \label{eq:appofK}
K(\sqrt{1 - \Psi ^2 (\a ) } ) \approx 
\ln \frac{4}{\Psi (\a ) }  
\end{equation}
and obtain
\begin{equation} \label{eq:scale}
a_1  a_2 \sim e^{-t} .
\end{equation}
Hence the smallest spatial scale for the passive scalar will decrease
exponentially in time.

In this paper, it is found that hyperbolic stagnation points can be a key
in the analysis the width of the Lagrangian analyticity strip of the
steady solution (\ref{psicellular}) to the 2-D Euler equation. 
Currently we are investigating the Lagrangian $\delta_\L(t)$
of the 3-D steady solutions (the ABC flows and more general Beltrami
flows \citep{MajdaBertozzi02}) of the 3-D Euler equations.
Since such flows can be chaotic, an interesting question is
which stretching --- caused by the chaos or the hyperbolic stagnation points
--- dominates. A preliminary result indicates
that the nearest singularities are again above the hyperbolic stagnation
points at large times. The details of the study of the 3-D steady flows
will be reported elsewhere.
It is also of interest to point out that
for kinematic dynamos in three-dimensional\footnote{Because of the 2-D 
antidynamo theorem three dimensions are required.}
steady flows, the fastest
growth of the magnetic field is frequently observed near hyperbolic
stagnation points \citep{soward1994} despite the presence of the chaotic
stretching.

A question for further study remains whether the behaviour of the Lagrangian
$\delta _\L (t) $ does exhibit some structural stability with respect 
to hyperbolic stagnation points. 

\vspace*{2mm}
\par\noindent {\bf Acknowledgments}.
We are grateful to J.~Bec and U.~Frisch for useful discussions.
Computational
resources were provided by the Yukawa Institute (Kyoto). This research was
supported by the European Union under contract HPRN-CT-2000-00162 and by the
Indo-French Centre for the Promotion of Advanced Research (IFCPAR~2404-2). 
WP would like to thank U.~Frisch, A.~Degenhard and Yu.~G.~Kondratiev 
for their support and help. 
TM was supported by the Japanese Ministry of Education Grant-in-Aid for Young
Scientists [(B), 15740237, 2003] and received also partial support from the
French Ministry of Education.

\section*{Appendix} 
\appendix
\renewcommand{\theequation}{\thesection.\arabic{equation}}
\section{Explicit construction of the Lagrangian map}
\setcounter{equation}{0}
\label{a:explicit}

The system of ordinary differential equations (\ref{eq 3}) can be solved
by a simple adaptation of what is done for an integrable case of the
ABC flow in \citep[Appendix A]{DFGHMS}.
For completeness we
give the full derivation, which is quite elementary.

Taking the derivative of (\ref{eq 3}) with respect to time we obtain a system of 
two decoupled differential equations
\begin{equation}
\left\{ \begin{array}{ll}
2 \ddot{x} _{1} = \sin 2 x_{1}  \\
2 \ddot{x} _{2} = -\sin 2 x_{2}, 
\end{array}
\right.
\label{eqmotionrewritten}
\end{equation}
satisfying the initial conditions (\ref{eq 4}) and 
\begin{equation} \label{addinitialconds}
\left\{ \begin{array}{ll}  \dot{x} _{1} (\a,\, 0)  = -\sin a_{1} \sin a_{2} \\ 
\dot{x} _{2} (\a,\, 0)  = - \cos a_{1} \cos a_{2} .
\end{array}
\right.
\end{equation}
Eq.~(\ref{eqmotionrewritten}) is obviously the same as a set of 
pendulum equations describing nonlinear oscillations around the origin 
for the $x_2$-variable and around $\pi/2$ for the $x_1$-variable.
Let us just consider the $x_2$-variable. The equation has the 
first integral \begin{equation} \label{firstintegral}
\frac{1}{2} \dot{x} _{2}^{2} - \frac{1}{4} \cos 2x_{2} = 
\frac{1}{4} - \frac{1}{2} \sin ^{2} a_{1} \cos ^{2} a_{2} =
\frac{1}{4} \left(
1-2 \Psi ^{2} (\mathbf{a} ) \right) ,
\end{equation}
which expresses the conservation of energy. From this equation 
we obtain by standard integration \citep{Lawden}
\begin{equation} \label{eq:solution}
\sin x_{2} (\mathbf{a} , t) = \sqrt{1-\Psi ^{2} (\mathbf{a} )} 
\mathrm{sn} \left( t + \int_{0}^{a_{2} } \frac{d\tilde{x} }{ \sqrt{(1-\Psi ^{2}
    (\mathbf{a} )) - \sin ^{2} \tilde{x} }} , \sqrt{1-\Psi^2 (\mathbf{a} ) } 
\right)
\end{equation}
Obviously, since the poles of the $\mathrm{sn} $ function do not lie on the 
real axis \citep{Lawden,AS65}, 
solutions of (\ref{eq 3}) are  
periodic and nonsingular for real initial conditions. The only possibility for 
an orbit to come across a pole and thereby to go to infinity 
is to start at a suitable complex location.

\end{document}